# Reactor for boron fusion with picosecond ultrahigh power laser pulses and ultrahigh magnetic field trapping

G H Miley[1], H Hora*[2,3] and G Kirchhoff[3]

[1]Deptartment of Nuclear, Plasma & Radiological Engineering, University of Illinois, Urbana IL, USA,
[2]Deptartment of Theoretical Physics, University of New South Wales, Sydney 2052, Australia
[3]UJK Management GmbH, Poing, Germany
*h.hora@unsw.edu.au

**Abstract.** Compared with the deuterium tritium (DT) fusion, the environmentally clean fusion of protons with $^{11}$B is extremely difficult. When instead of nanosecond laser pulses for thermal-ablating driven ignition, picoseconds pulses are used, a drastic change by nonlinearity results in ultrahigh acceleration of plasma blocks. This radically changes to economic boron fusion by a measured new avalanche ignition.

## 1. Introduction

In search for the aim of energy production by the nuclear fusion of light nuclei, an alternative may be available by using laser pulses of picosecond (ps) duration and powers in the range above dozens of petawatts (PW) up to terawatts (TW). The essential difference to the highly developed laser fusion by using huge lasers with nanosecond laser pulses of more than MJ energy [1] now with the achievements of "high foot" operation [2] is that the non-thermal forces of laser pulses are used in combination with recently achieved ultrahigh magnetic fields for trapping the reacting plasma. In addition use is done by the avalanche reaction of fusing light hydrogen (protons) with the boron isotope $^{11}$B (HB11) with negligible problems of radioactive pollution [3][4] and direct conversion of the of the energy of the generated alpha particles into electric power [5][6].

This option was developed from the physics of the basically nonlinear processes opened by the interaction of lasers with plasmas nearly from the very beginning with the special turning point [7] of generating the >PW powers by Chirped Pulse Amplification CPA [8]. The ultrahigh acceleration was calculated since 1978 ([9]: Figs. 10.17 a&b) by direct conversion of laser energy into mechanical motion of macroscopic plasma bocks by nonlinear (ponderomotive) forces and first measured [10] in full agreement with the theory [11]. The necessary picosecond interaction time for igniting uncompressed solid density deuterium-tritium (DT) fusion fuel [12] could be shown to be possible for the otherwise extremely difficult clean fusion of HB11 [13][14]. The need of the ultrahigh magnetic fields [15] and of the avalanche reaction of HB11 [3][4] was the final step for this development.

In discussing this option [3], Crandall [6] mentioned that an enormous further work would be necessary to achieve this goal. The following views may summarize some of the tasks.

## 2. Laser development

The building of lasers of higher powers is exciting for all kinds of applications and sub-picosecond lasers have achieved 7.2 PW pulse powers [16]. New laser architectures based on existing high energy PW technologies [8] for production of 200 PW laser pulses from single beam lines of existing NIF and NIF-like lasers have been recently defined [17]. Laser pulses of 200 PW are on the way by using the existing NIF laser. NIF-like lasers can be reduced in size by many orders of magnitude after short laser-diode-pumped neodymium glass laser amplifiers are ready for use.

For the ps-block ignition of boron fusion, very high quality laser pulses with extreme contrast are necessary as seen from the repetition [18] of the Sauerbrey's [10] ultrahigh acceleration in agreement with the theory. This technology can be studied immedietly with table top laser experiments for generating plasma blocks with several 100 MeV ion energy having space charge neutral ion densities million times higher than from accelerators. This is needed also for other applications as e.g. for hadron therapy of cancer [19].

Plasma mirror techniques can evaluate to what degree the contrast is necessary where it can be explored what initial density profiles of the irradiated targets should be used for optimized conditions of the nonlinear force driven dielectric explosion [4][9]. It had been shown numerically (see Figure 1 of [3]) that the use of a bi-Rayleigh initial density profile for a very low reflection of the laser light has the additional advantage that the plasma blocks for acceleration can have an increased thickness for the dielectric explosion. Evaluations with modified density variations of the target can strongly increase the energy of the ions in the accelerated plasma bocks [19]. Numerical studies for providing optimized targets e.g. as foam with a varying density profile can experimentally confirm the expected optimization for HB11 fusion.

## 3. Ultrahigh magnetic fields and the avalanche reaction

The measured ultrahigh magnetic fields of 4.5 kilotesla for about nanosecond by laser pulses with kJ energy of nanosecond duration [15] can be used to trap a cylindric plasma from solid density HB11. The generated fluid of alpha particles from the 0.2mm thin target expands slowly in radial direction (see [21]: Fig. 3) but with many1000km/s along the cylinder axis when trapped by 10 kT fields even when computing very pessimistically used binary fusion reactions, well with high gains of ignition [21]. When including the avalanche reaction and the subsequent expansion into a solid density mantle of 1mm radius around the initial 0.2mm cylinder, the block ignition by a ps laser pulse of 30kJ energy should produce more than GJ energy in alpha particles. Experimental studies of this generation of ultrahigh magnetic fields should be studied specifically to measure the trapping of the co-axial, initially solid HB11 fuel when compressed by the magnetic field in comparison to preceding and further generalized computations with cylindrical rotation [21][22][23][24].

Independent from these steps, the measurement of $10^9$ HB11 reactions per steradian [25] at laser plane geometry reaction per laser irradiation on very special targets was perfectly repeated [26]. This is the basic measurement for concluding the avalanche process [3][4]. Further studies under varying conditions should well be performed to further confirm these very exceptionally increased HB11 reaction gains.

Then it may be interesting to combine these measurements with cylindrical HB11 targets in presence of the ultrahigh magnetic fields with measuring the generation of alpha particles. It can be expected that the reaction rate may be further increased above the values reached numerically [21] when the fact of the gyration of the alphas are added on top of the avalanche mechanism.

For further confirming the theory of the avalanche process, the elastic ion collisions in the broad energy range of 600keV energy will be further evaluated. This is a state of a highly non-equilibrium plasma. A number of results for describing these conditions are at hand in support of the avalanche reactions [27] for the centre of mass distribution in the reacting plasma. This can be combined with the results of multi-fluid computations where the generated fluid of alpha particles [21] resulted in the ignition process even for the pessimistic HB11 binary reactions without avalanche. This led to realize that the reacting plasma was in an extreme non-equilibrium state, and it was possible to use the elastic collisions of the alphas with protons and boron nuclei to reach the high fusion gains.

These results of the elastic nuclear interaction for the protons and boron nuclei at the very broad energy distribution with a maximum of ion interaction energies at 600 keV is exceptional for HB11 due to the fact that there is the very broad maximum of the energy dependence of the fusion reaction cross sections. HB11 has - in contrast to all other fusion reactions - a cross section very broadly elevated by a factor of more than ten for reproducing [27] the measured strong increase of the HB11 reaction gains [25].

## 4. Scheme for designing of a HB11 reactor

The numerical result of generation of more than one gigajoule energy (270 kWh) of alpha particles from a solid HB11 cylinder of 1 cm length and 1mm radius in a co-axial 10 kilotesla magnetic field during one ns by irradiation of one ps laser pulse of 30 kJ at one side at $10^{20}$ W/cm$^2$ [4][21] should convert at least the main part of the generated alpha particle energy into electric power by being slowed down electrostatically in a field of 1.4 megavolt (Figure 1) [4][5].

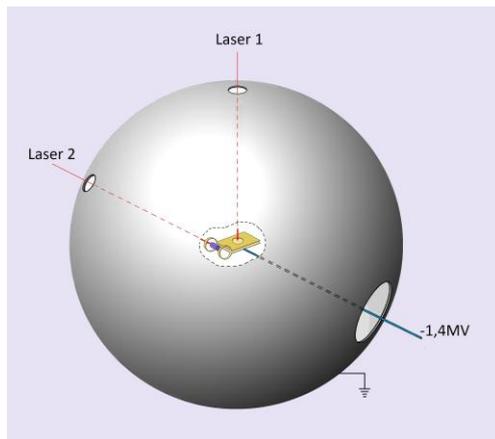

*Figure 1. Scheme of a HB11 fusion reactor without any radioactive radiation problem is based on non-thermal plasma block ignition by nonlinear forces using a 30kJ-picosecond laser pulse 2. The central reaction unit (Figure 2 of [3]) located in the center of the reactor sphere is electric charged to the level of -1.4 million volts against the wall of the sphere such that the alpha particles (helium nuclei) produce more than a gigajoule energy, of which a small part is needed for the operation of the laser pulses [3][4][5].*

The reaction in the center of the sphere with producing more than GJ energy generates a mechanical shock to the spherical container of the reactor. This is to be compared with the shock of chemical explosives. The generated momentum of the alpha particles compared with a chemical explosion depends on the square root of the energy of the particles which is for the nearly 10-Million times higher nuclear than that of chemical energy. The mechanical shock of the alpha particles to the reactor wall is then about 3000 times lower than for a comparable chemical explosive. In the case of the reactor of Figure 1, this corresponds then to a number of several grams of TNT.

It can be clarified that the boron fusion reaction occurring only in the volume of few cubic millimetres and during the time of less than a few nanoseconds is limited to the controlled generation of fusion energy and cannot be used for large scale nuclear reactions.

The technology for handling the loading of the central HB11 reaction unit into the sphere in Figure 1 within vacuum and on an electric potential of -1.4 megavolts, especially when operating with one reaction per second, needs a special design of the equipment to be used, but this should be on a level of present day's achievements. Special attention may have to be given to move the unit through gates from zero potential to the high potential in the interior of the reactor to be located at the right hand side of Figure 1.

If the reactor has to run with one shot per second, the direct current electrical energy in the pulse has to be converted according to the methods of the High Voltage Direct Current HVDC techniques in three phase ac current. This technology is completely developed for commercial use of power transmission over large distances [28][29][30] such that the generated power at the one Hertz operation corresponds to a 1000 Megawatt power station.

A very preliminary first estimation after subtracting of the costs for the central reaction unit in the center of the sphere of Figure 1 which is destroyed at each shot and of the costs for the HB11 fuel, and after subtracting the costs of the investment and that of the operation of the power station, the profit may arrive at up to $300Million per year of generated electric power grid.